\documentclass[journal=jctcce,manuscript=article]{achemso}
\setkeys{acs}{articletitle=true}
\usepackage{graphicx}
\usepackage{amsmath}
\usepackage{amsfonts}
\usepackage{amssymb}
\usepackage{amsthm}
\usepackage{braket}
\usepackage{color}
\usepackage{achemso}
\usepackage{bm}
\usepackage{booktabs}
\usepackage{comment}
\usepackage{multirow}
\usepackage{upgreek}
\usepackage{hyperref}
\usepackage{xspace}
\usepackage[nolist]{acronym}
\usepackage{siunitx}

\usepackage[utf8x]{inputenc}

\begin{acronym}
\acro{API}{application programming interface}
\acro{QM/MM}{quantum mechanics/molecular mechanics}
\acro{DC}{dielectric continuum}
\acro{GEPOL}{GEPOL}
\acro{IEF}{Integral Equation Formalism}
\acro{LHS}{left-hand side}
\acro{MD}{Molecular Dynamics}
\acro{MM}{molecular mechanics}
\acro{PCM}{polarizable continuum model}
\acro{BVP}{boundary value problem}
\acro{BIE}{boundary integral equation}
\acro{PDE}{partial differential equation}
\acro{PE}{polarizable embedding}
\acro{QC}{Quantum Chemistry}
\acro{RHS}{right-hand side}
\acro{DFT}{density functional theory}
\acro{SAS}{Solvent-Accessible Surface}
\acro{SES}{Solvent-Excluded Surface}
\acro{ASC}{apparent surface charge}
\acro{MEP}{molecular electrostatic potential}
\acro{SCF}{self-consistent field}
\acro{MO}{molecular orbital}
\acro{TDSCF}{time-dependent SCF}
\acro{KS}{Kohn--Sham}
\acro{XC}{exchange--correlation}
\acro{OPA}{one-photon absorption}
\acro{TPA}{two-photon absorption}
\acro{MPA}{multiphoton absorption}
\acro{PNA}{\emph{para}-nitroaniline}
\acro{PDNB}{\emph{para}-dinitrobenzene}
\acro{MCP}{methylenecyclopropene}
\acro{QM}{quantum-mechanical}
\acro{VEE}{vertical excitation energy}
\acrodefplural{VEE}[VEEs]{vertical excitation energies}
\acro{FQ}{fluctuating charge}
\acro{R6G}{rhodamine 6G}
\acro{EEP}{electronegativity equalization principle}
\acro{FF}{force field}
\acro{AO}{atomic orbital}
\acro{SI}{supporting information}
\end{acronym}

\newcommand*{\diff}{\mathop{}\!\mathrm{d}} 

\newcommand*{\pderiv}[3][]{%
\frac{\partial^{#1}#2}
{\partial #3^{#1}}}
\newcommand*{\vect}[1]{\bm{#1}} 
\newcommand*{\mat}[1]{\bm{#1}}
\newcommand*{\Tr}{\mathrm{Tr}}

\newcommand{\hairsp}{\hspace{1pt}}
\newcommand{\ie}{\textit{i.\hairsp{}e.}\xspace}

\newcommand{\eg}{\textit{e.\hairsp{}g.}\xspace}

\newcommand*{\diels}{\varepsilon_\mathrm{s}}
\newcommand*{\dield}{\varepsilon_\infty}

\newcommand*{\pertmat}[1]{\tilde{\mat{#1}}}
\newcommand*{\aveL}{\lbrace\tilde{L}(\pertmat{C}, \pertmat{\mu}, \pertmat{q}, t)\rbrace_T}
\newcommand*{\aveLa}{\lbrace\tilde{L}^a(\pertmat{D}, \pertmat{q}, t)\rbrace_T}

\newcommand*{\eqtr}{\overset{\Tr}{=}}
\newcommand*{\eqavetr}{\overset{\lbrace\Tr\rbrace_T}{=}}

\newcommand*{\aherm}[1]{\left[#1\right]^{\ominus}}

\newcommand*{\rspParam}[1]{\mat{X}^{#1}}
\newcommand*{\RHS}[1]{\mat{M}_\mathrm{RHS}^{#1}}
\newcommand*{\Dp}[1]{\mat{D}_\mathrm{P}^{#1}}

\newcommand*{\genHessian}{\mat{E}^{[2]}}
\newcommand*{\genMetric}{\mat{S}^{[2]}}

\newcommand*{\mathcolor}{}
\def\mathcolor#1#{\mathcoloraux{#1}}
\newcommand*{\mathcoloraux}[3]{%
  \protect\leavevmode
    \begingroup
        \color#1{#2}#3%
          \endgroup
}

\graphicspath{{img/}} 

\author{Roberto Di Remigio$^\top$}
\email{roberto.d.remigio@uit.no}
 \affiliation{Hylleraas Centre for Quantum Molecular Sciences, Department of Chemistry, University of Troms{\o} - The Arctic University of Norway, N-9037 Troms{\o}, Norway}
 \alsoaffiliation{Department of Chemistry, Virginia Tech, Blacksburg, Virginia 24061, United States}
\author{Tommaso Giovannini$^\top$}
\email{tommaso.giovannini@ntnu.no}
\affiliation{Department of Chemistry, Norwegian University of Science and Technology, 7491 Trondheim, Norway}
\author{Matteo Ambrosetti}
\affiliation{Scuola Normale Superiore, Piazza dei Cavalieri 7, 56126 Pisa, Italy}
\author{Chiara Cappelli}
\affiliation{Scuola Normale Superiore, Piazza dei Cavalieri 7, 56126 Pisa, Italy}
\author{Luca Frediani}
\affiliation{Hylleraas Centre for Quantum Molecular Sciences, Department of Chemistry, University of Tromsø - The Arctic University of Norway, N-9037 Tromsø, Norway}

\abbreviations{TPA, FQ, QM/MM}
\keywords{Hybrid models, multiphoton absorption, response theory}

\title[TPA-FQ]{Fully Polarizable QM/Fluctuating Charge Approach to Two-Photon Absorption of Aqueous Solutions}

\begin{document}

\begin{center}
$^\top$These authors contributed equally to this work
\end{center}

\begin{abstract}
We present the extension of the quantum/classical polarizable \acl{FQ} model
to the calculation of single residues of quadratic response functions, as
required for the computational modeling of two-photon absorption cross-sections.
By virtue of a variational formulation of the quantum/classical polarizable
coupling, we are able to exploit an atomic orbital-based quasienergy formalism
to derive the additional coupling terms in the response equations.
Our formalism can be extended to the calculation of arbitrary order response
functions and their residues.
The approach has been applied to the challenging problem of one- and
two-photon spectra of \ac{R6G} in aqueous solution.
Solvent effects on one- and two-photon spectra of \ac{R6G} in aqueous
solution have been analyzed by considering three different approaches, from a continuum (QM/PCM) to two QM/MM models (non-polarizable QM/TIP3P and
polarizable QM/FQ). Both QM/TIP3P and QM/FQ simulated OPA and TPA spectra show
that the inclusion of discrete water solvent molecules is essential to increase
the agreement between theory and experiment. QM/FQ has been shown to give the best agreement with experiments.
\end{abstract}

\section{Introduction}\label{sec:intro}

Multiphoton absorption is the synchronous absorption of multiple photons leading
to an excitation of a molecule from one electronic state to
another.\cite{he2008multiphoton} Such an effect was originally predicted by
G{\"o}ppert-Mayer in 1931,\cite{goppert1931elementarakte} but only first measured in 1961 because its intensity was too weak to be detected before the advent of laser sources.\cite{kaiser1961two}
Among multiphoton processes, the most common is \ac{TPA}, in which the
simultaneous absorption of two photons takes place.\cite{he2008multiphoton}
Nowadays, \ac{TPA} measurements are not as common as compared to \ac{OPA},
however the study of \ac{TPA} processes is growing and rapidly becoming a
well-established research field.\cite{terenziani2008enhanced,pawlicki2009two}
\ac{TPA} is governed by different symmetry selection rules, states that are dark
in \ac{OPA} experiments might thus be accessible in \ac{TPA}.
Furthermore, \ac{TPA} is a non-linear process whose intensity depends on the square of the incoming light. This affords a greater spatial resolution than in \ac{OPA} experiments. \ac{TPA} has a number of technological applications, in particular in molecular devices.\cite{he2008multiphoton,parthenopoulos1989three,
  gorman2004vitro,kohler1997exchange,belfield2000near}
The design of molecular systems with large \ac{TPA} cross sections is thus a challenge both from an experimental and computational point of view.\cite{beerepoot2015benchmarking,friese2012calculation,guillaume2010computational,
toro2010two,ferrighi2007two,steindal2012combined,
nag2009solvent,paterson2006benchmarking,hattig1998multiphoton,he2008multiphoton,
albota1653design,nanda2015two,woo2005two,woo2005water,woo2005solvent,abbotto2002novel}

High TPA cross sections are generally measured for large
chromophores,\cite{albota1653design,chakrabarti2009large} for which the
computational description at high level of accuracy is difficult and sometimes
not affordable. For this reason, most of the computational studies on this kind
of spectroscopy have been performed by resorting to \ac{DFT}, due to the good
compromise between accuracy and computational
cost.\cite{beerepoot2015benchmarking,beerepoot2018benchmarking,salek2003calculations}
In addition to the quantum-mechanical issue in the description of the target
molecule, it is worth remarking that most of the experimental TPA cross sections
are measured in the condensed
phase.\cite{woo2005two,woo2005water,woo2005solvent,kato2006novel,terenziani2006linear}
For instance, a 50\% increase in TPA cross sections has been reported by changing the solvent.\cite{terenziani2008enhanced}
In order to successfully reproduce experimental data, such effects need to be taken into consideration.

The problem of treating solvent effects on observable properties is one of the
pillars in quantum chemistry. The most successful approaches make use of multiscale and focused
models, where the environment is treated at a lower level of accuracy with respect to the
target molecule:
\cite{warshel1976theoretical,warshel1972calculation,tomasi2005,senn2009qm,mennucci2013modeling,
cappelli2011towards,egidi2014stereoelectronic,boulanger2018qm,
lahiri2014large,loco2018modeling,diremigio2017four}
the latter is generally treated at the \ac{QM} level,
whereas the former is treated classically.

In the resulting QM/classical approaches, the classical portion can range from
an atomistic description (giving rise to \ac{QM/MM}
models\cite{warshel1976theoretical,warshel2003computer,warshel1972calculation,
Gao1992,curutchet2009electronic,Gordon_JPCA_EFP,Barone_Libro_QMMM,nielsen2007density})
to a \ac{DC} description. Among the latter, the
\ac{PCM},\cite{miertuvs1981electrostatic,tomasi1994molecular,orozco2000theoretical,tomasi2005,
mennucci2007continuum,scalmani2010continuous, lipparini.2010} in which the
environment is depicted as a homogeneous continuum dielectric with given
dielectric properties, has been particularly successful.
In such an approach, the QM described target molecule is accommodated into a
molecular shaped cavity. The QM electron density and the dielectric mutually
polarize. The QM/PCM approach has been extended to the description of TPA spectra by
some of the present authors,\cite{frediani2005two,zhao2007solvent,frediani2005second,ferrighi2007two}
and an open-ended response formulation was put forth in a recent
communication.\cite{diremigio2017open} One of the main problems related to a
continuum description of the environment is that all information about the
atomistic structure of the environment is neglected. Thus, the specific
molecule-environment interactions (\eg hydrogen bonding), cannot be described.

In order to recover the atomistic description of the environment, QM/MM is
exploited, where the target molecule is still described at the \ac{QM} level, whereas
the environment is described by resorting to \ac{MM} force
fields.\cite{warshel1976theoretical,field1990combined,gao1996hybrid,friesner2005ab,lin2007qm,
senn2009qm,monari2012theoretical,monari2012qm,boulanger2018qm}
In electrostatic QM/MM embedding approaches, a set of fixed charges is placed on
the MM portion and the interaction between QM density and MM charges is
introduced in the QM Hamiltonian.
Mutual polarization, \ie{} the polarization of the MM portion arising from the
interaction with the QM density and viceversa, can be introduced by employing
polarizable force fields.
These can be based on distributed
multipoles,\cite{day1996effective,kairys2000qm,mao2016assessing,loco2016qm,loco2018modelingijqc}
induced
dipoles,\cite{thole1981molecular,steindal2011excitation,jurinovich2014fenna}
Drude oscillators,\cite{boulanger2012solvent} capacitances and
polarizabilities\cite{Rinkevicius2014-ez,Rinkevicius2014-by,Li2014-gq},
\acp{FQ}\cite{rick1994dynamical,rick1996dynamical,cappelli2016integrated} or fluctuating charges and fluctuating dipoles (FQF$\mu$).\cite{giovannini2019fqfmu}

Thanks to the availability of a variational formulation of the quantum/classical
polarizable coupling, the QM/FQ approach has been extended to the
analytical calculation of a large variety of properties and spectroscopies:
molecular gradients
and Hessians,\cite{lipparini2012analytical} linear response
properties,\cite{lipparini2012linear,giovannini2018simulating} including optical
rotation,\cite{lipparini2013optical,egidi2015optical,egidi2019combined} and
electronic circular dichroism,\cite{egidi2015electronic} vibrational circular
dichroism,\cite{giovannini2016effective} third order mixed
electric/magnetic/geometric
properties\cite{giovannini2017polarizable,giovannini2018effective} and second
harmonic generation.\cite{giovannini2018hyper} Remarkably, QM/FQ has already been shown to accurately model some of the systems where PCM and other continuum models completely fail.

Non-polarizable QM/MM approaches and polarizable QM/MM based on induced dipoles
have been extended to the calculation of TPA spectra of molecules in
solution.\cite{nielsen2007density,steindal2011excitation,steindal2016open} In
this paper, we extend the QM/FQ model to the computation of \ac{TPA} spectra.
In particular, we have selected the challenging case of \acf{R6G} in
aqueous solution, which has been studied extensively both theoretically and
experimentally.\cite{makarov2008two,milojevich2013surface,milojevich2011probing,guthmuller2008resonance1,
guthmuller2008resonance2,dieringer2008surface,weiss2014non,nag2009solvent} The
large interest in such a molecule -- in particular to its TPA spectrum -- is due
to the transition between the ground and second excited state, which is
dark in OPA due to symmetry selection rules.
From a theoretical point of view, this is the first time that solvent
effects on TPA of \ac{R6G} in aqueous solution are considered. This is
achieved using a variety of models: a continuum approach (PCM), an electrostatic embedding
(QM/TIP3P\cite{mark2001structure}) and a polarizable embedding (QM/FQ).

The paper is organized as follows: we first describe the QM/FQ approach and
derive its extension to quadratic response properties using a quasienergy
formulation. After briefly discussing our implementation and the computational
protocol, we discuss our \ac{OPA} and \ac{TPA} results for the \ac{R6G} system
in aqueous solution.
Except where stated otherwise, atomic units are used throughout.

\section{Theory and Implementation}\label{sec:theory}

\subsection{The QM/FQ Approach}

In the FQ approach, each MM atom is endowed with a charge which can vary
according to the \ac{EEP}\cite{Mortier1985,sanderson1951} which states that, at equilibrium, the
instantaneous electronegativity ($\chi$) of each atom has the same
value.\cite{sanderson1951,Mortier1985}
The model is based on a set of two parameters, \ie{} atomic electronegativies
and chemical hardnesses ($\eta$), which can be rigorously defined within
\emph{conceptual DFT}\cite{Mortier1985,chelli.2002} as the first and second
derivatives of the energy with respect to the charges, respectively.
Through these parameters, \acp{FQ} ($q$) can be defined as those minimizing the
functional:\cite{lipparini2011polarizable,rick1994dynamical}
\begin{equation}
\label{eq:Func}
\begin{aligned}
  F[\mat{q},\boldsymbol{\lambda}]
  &=
\sum_{\alpha,i} q_{\alpha i}\chi_{\alpha i} + \frac{1}{2}\sum_{\alpha,i} \sum_{\beta, j}q_{\alpha i} J_{\alpha i,\beta j}q_{\beta j} + \sum_{\alpha}\lambda_{\alpha} \left(\sum_{i} q_{\alpha i} - Q_{\alpha}\right) \\
  &= \mat{q}^\dagger\boldsymbol{\chi} + \frac{1}{2}\mat{q}^\dagger\mat{J}\mat{q} + \mat{q}^{\dagger}\boldsymbol{\lambda}
\end{aligned}
\end{equation}
where $\mat{q}$ is a vector containing the FQs, the Greek indices $\alpha$
run over molecules and the Latin ones $i$ over the atoms of each molecule.
$\boldsymbol{\lambda}$ is a set of Lagrangian multipliers used to impose charge
conservation constraints on each molecule. The charge interaction kernel
$\mat{J}$ is, in our implementation, the Ohno kernel and
the diagonal terms of $\mat{J}$ kernel are the chemical hardnesses $\eta$.
The stationarity conditions of the functional in eq.\eqref{eq:Func} are defined through a linear system:\cite{giovannini2018hyper,lipparini2011polarizable}
\begin{equation}
  \label{eq:sysqeq}
  \begin{pmatrix}
    \mat{J} & \mat{1}_{\boldsymbol{\lambda}} \\
    \mat{1}^{\dagger}_{\boldsymbol{\lambda}} & \mat{0} \\
  \end{pmatrix}
  \begin{pmatrix}
    \mat{q} \\
    \boldsymbol{\lambda}
  \end{pmatrix}
=
-
  \begin{pmatrix}
    \boldsymbol{\chi} \\
    \mat{Q}
  \end{pmatrix}
\end{equation}
We note in passing that the capacitance \ac{MM} model employed by
\citeauthor{Rinkevicius2014-by} includes induced charges in its modeling of the
metallic portions of the \ac{MM} environment.\cite{Rinkevicius2014-by} Hence,
despite the largely different physical setting of the models, their polarization
equations bear a significant resemblance.

\begin{figure}
\centering
\includegraphics[width=.4\textwidth]{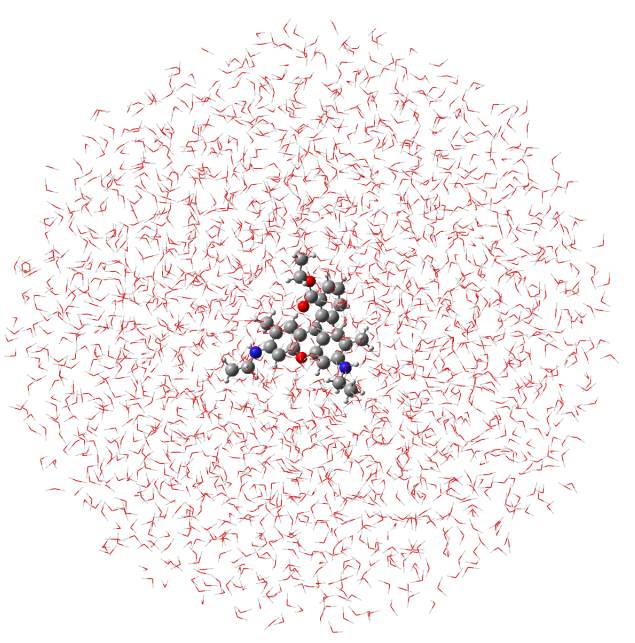}
\caption{Schematic representation of the QM/FQ model.}
\label{fig:pallaQMFQ}
\end{figure}

The QM/FQ model system is constituted by a QM core region placed at the center
of a spherical region defining the environment (see Figure \ref{fig:pallaQMFQ}),
\ie containing a number of solvent molecules, which are described in terms of \ac{FQ} \ac{FF}.
The FQ \ac{FF} can be effectively coupled to \emph{any} QM method. It suffices
to augment the QM energy functional with the classical functional in
eq.~\eqref{eq:Func} and the quantum/classical interaction.
For the FQ \ac{FF} the latter is defined as the classical electrostatic interaction between the \acp{FQ} and the QM density.\cite{lipparini2012linear}
Explicitly, the interaction takes the form:
\begin{equation}
 \label{eq:EInt}
 E_{\mathrm{QM/FQ}}
 =
 \sum_{i=1}^{N_{\mathrm{FQ}}} V[\rho](\mathbf{r}_i)q_i
 =
 \sum_{i=1}^{N_{\mathrm{FQ}}} \sum_{\mu,\nu = 1}^{N_{\mathrm{AO}}}\Braket{\chi_{\mu} | \frac{-q_{i}}{\left|\mathbf{r} - \mathbf{r}_{i}\right|} | \chi_{\nu}}D_{\nu\mu},
\end{equation}
with $V[\rho](\mathbf{r}_i)$ the electrostatic potential due to
the QM density of charge at the $i$-th FQ $q_i$ placed at $\mathbf{r}_i$,
$\lbrace \chi_{\mu}\rbrace$ a Gaussian \ac{AO} basis set and $D_{\mu\nu}$ the
\ac{AO} density matrix.
For a hybrid \ac{KS} \ac{DFT} description of the \ac{QM} moiety, the global QM/MM energy functional reads:\cite{lipparini2012linear,lipparini2012analytical}
\begin{equation}
\label{eq:1-FQSCF}
\mathcal{E}[\mat{D},\mat{q},\boldsymbol{\mat{\lambda}}]
=
E_{\mathrm{SCF}}[\mat{D}] + F[\mat{q}, \boldsymbol{\lambda}] + \mat{q}^\dagger\mat{V}(\mat{D}).
\end{equation}
Stationarity of eq.~\eqref{eq:1-FQSCF} with respect to the density matrix yields
the \ac{SCF} equations:
\begin{equation}
  \mathbf{F}\mat{C} = \mat{S}\mat{C}\boldsymbol{\epsilon},\quad
  \mat{F} = \mat{h} + \mat{G}^{\gamma}(\mat{D}) + \mat{F}_{\mathrm{xc}} + \mat{F}_{\mathrm{FQ}}
\end{equation}
where the various terms in the \ac{KS} matrix $\mat{F}$ are:
\begin{subequations}
  \begin{align}
    h_{\mu\nu} &=\Braket{ \chi_\mu |
    -\frac{1}{2}\nabla^2
    - \sum_K\frac{Z_K}{\left|\vect{R}_K - \vect{r}\right|} |
    \chi_\nu } \text{,} \label{eq:h-matrix}\\
    G_{\mu\nu}^\gamma(\mat{D}) &=
    \sum_{\alpha\beta}D_{\beta\alpha}(g_{\mu\nu\alpha\beta}
                                    - \gamma g_{\mu\beta\alpha\nu}) \text{,} \label{eq:G-matrix} \\
  F_{\mathrm{xc}, \mu\nu} &=
    \int
    \diff \vect{r}
    \chi_{\mu}(\vect{r})\chi_{\nu}(\vect{r})\left.\pderiv{E_\mathrm{xc}}{\rho(\vect{r})}\right|_{\rho(\vect{r})}
    =
    \int
    \diff \vect{r}
    \chi_{\mu}(\vect{r})\chi_{\nu}(\vect{r})v_{\mathrm{xc}}(\vect{r})\text{,} \label{eq:XC-matrix} \\
 F_{\mathrm{FQ}, \mu\nu} &= \mat{q}^{\dagger}\mat{V}_{\mu\nu} = \sum_{i = 1}^{N_{\mathrm{FQ}}} \Braket{\chi_{\mu} | \frac{-q_{i}}{\left|\mathbf{r} - \mathbf{r}_{i}\right|} | \chi_{\nu}}\text{.}
 \label{eq:FockQMFQ}
  \end{align}
\end{subequations}
The \acp{FQ} consistent with the QM density are obtained by solving the
stationarity conditions with respect to the polarization variational degrees of
freedom $\mat{q}$, \ie by solving eq.~\eqref{eq:sysqeq} with a modified \ac{RHS}, including the QM potential as additional source term, effectively coupling the QM and MM moieties and ensuring mutual polarization:
\begin{equation}
 \label{eq:FQQM}
  \begin{pmatrix}
    \mat{J} & \mat{1}_{\boldsymbol{\lambda}} \\
    \mat{1}^{\dagger}_{\boldsymbol{\lambda}} & \mat{0} \\
  \end{pmatrix}
  \begin{pmatrix}
    \mat{q} \\
    \boldsymbol{\lambda}
  \end{pmatrix}
=
-
  \begin{pmatrix}
    \boldsymbol{\chi} + \mat{V}(\mat{D}) \\
    \mat{Q}
  \end{pmatrix}
\end{equation}.

\subsubsection{Linear and Quadratic Response Functions in a QM/FQ framework}

Thanks to its variational formalism, the QM/FQ approach\cite{cappelli2016integrated} is especially suited to the
modeling of response and spectral properties because its energy expression can
be easily differentiated up to high orders.
The quantum/classical coupling terms needed for the calculation of response
properties, can be easily derived and implemented so that
polarization effects are fully considered also in the computed final spectral
data.\cite{lipparini2012linear,lipparini2012analytical,lipparini2013gauge,giovannini2016effective,
giovannini2017polarizable,giovannini2018simulating,carnimeo2015analytical}

For a QM/FQ system subject to a Hermitian, time-periodic, one-electron
perturbation $V^{t}$, response functions and response equations can be
formulated in the atomic orbital, density matrix-based quasienergy formalism\cite{helgaker2012recent} of
\citet{Thorvaldsen2008-sg}. To the best of our knowledge, this is the first time
the quasienergy formalism is employed in the QM/FQ framework.
The starting point is the time-averaged quasienergy Lagrangian, $\aveL$, parametrized in terms of the
desired perturbed coefficient matrix $\pertmat{C}$, the Lagrange multipliers
ensuring orthonormality of the one-electron basis $\pertmat{\mu}$ and the perturbed \acp{FQ}
$\pertmat{q}$. The tilde is here used for quantities evaluated at
general perturbation strengths.
We can obtain this Lagrangian by augmenting the quasienergy in
eq.~(52) of Ref.~\citenum{Thorvaldsen2008-sg} with the perturbed \ac{FQ} functional of eq.~\eqref{eq:Func}.
The time-averaged quasienergy Lagrangian is not suitable for an atomic
orbital-based theory, since it features the \ac{MO} coefficient matrix.
However, its \emph{perturbation-strength-differentiated} counterpart
$\aveLa$ can be expressed in terms of the desired variational degrees of
freedom $\pertmat{D}$ and $\pertmat{q}$:
\begin{equation}\label{eq:aveLa}
 \tilde{L}^a(\pertmat{D}, \pertmat{q}, t) \eqavetr \tilde{\mathcal{E}}^{00,a} - \pertmat{S}^a\pertmat{W},
\end{equation}
where we have borrowed notation from Ref.~\citenum{diremigio2017open}. Indeed,
the close similarity between the \ac{PCM} and \ac{FQ} models allows us to leverage
the same arguments in Ref.~\citenum{diremigio2017open} to formulate linear and
quadratic response functions in a QM/FQ framework.
The generalized \ac{KS} energy $\tilde{\mathcal{E}}$ is the time-dependent
equivalent of eq.~\eqref{eq:1-FQSCF}:
\begin{equation}
\label{eq:2-FQSCF}
\tilde{\mathcal{E}}[\pertmat{D}, \pertmat{q}]
 \eqtr
 \left[\pertmat{h} + \pertmat{V}^{t} +
 \frac{1}{2}\pertmat{G}^{\gamma}(\pertmat{D}) -
 \frac{\mathrm{i}}{2}\pertmat{T}\right]\pertmat{D}
 + \tilde{E}_\mathrm{xc}[\tilde{\rho}(\pertmat{D})] + h_\mathrm{nuc}
 + F[\pertmat{q}, \pertmat{\lambda}] + \pertmat{q}^\dagger\pertmat{V}(\pertmat{D}) \text{.}
\end{equation}
where $\pertmat{V}^{t}$ is the \ac{AO} basis representation of the perturbation
operator and $\pertmat{T} = \Braket{\tilde{\chi}_\mu|\dot{\tilde{\chi}}_\nu} -
\Braket{\dot{\tilde{\chi}}_\mu|\tilde{\chi}_\nu}$ is the time-differentiation
\ac{AO} overlap matrix.
Evaluation of eq.~\eqref{eq:aveLa} at zero perturbation strength yields the the
first-order property formula. For an electric field perturbation this
corresponds to the electric dipole moment and, thanks to the Hellmann--Feynman
theorem, only requires the unperturbed density matrix.
Further differentiation of eq.~\eqref{eq:aveLa} and evaluation at zero perturbation
strength yields higher order response functions. Detailed expressions can be
obtained from eqs.~(22a)-(22c) and Appendix A in
Ref.~\citenum{diremigio2017open} by replacing perturbed and unperturbed apparent
surface charges $\sigma$ with perturbed and unperturbed \acp{FQ}, respectively and
the generalized free energy $\mathcal{G}$ with $\mathcal{E}$:
\begin{subequations}
  \begin{align}
    L^{ab} \eqavetr &\mathcal{E}^{00,ab}
    + \mathcal{E}^{10,a}\mat{D}^{b}
    + \mathcal{E}^{01,a}\mat{q}^{b}
    - \mat{S}^{ab}\mat{W}
    - \mat{S}^{a}\mat{W}^{b} \label{eq:second-order} \\
    \begin{split}
    L^{abc} \eqavetr
      &\mathcal{E}^{00,abc}
    + \mathcal{E}^{10,ac}\mat{D}^{b}
    + \mathcal{E}^{10,ab}\mat{D}^{c}
    + \mathcal{E}^{20,a}\mat{D}^{b}\mat{D}^{c}
    + \mathcal{E}^{10,a}\mat{D}^{bc}
    + \mathcal{E}^{11,a}\mat{D}^{b}\mat{q}^c \\
    &+ \mathcal{E}^{01,ac}\mat{q}^{b}
    + \mathcal{E}^{01,ab}\mat{q}^{c}
    + \mathcal{E}^{01,a}\mat{q}^{bc}
    + \mathcal{E}^{11,a}\mat{q}^{b}\mat{D}^c \\
    &- \mat{S}^{abc}\mat{W}
    - \mat{S}^{ab}\mat{W}^{c}
    - \mat{S}^{ac}\mat{W}^{b}
    - \mat{S}^{a}\mat{W}^{bc}
    \end{split} \label{eq:third-order}
  \end{align}
\end{subequations}

Response parameters need to be determined in order to assemble the property
expressions from the perturbed variational parameters $\mat{D}^{a}$,
$\mat{q}^{a}$ and so forth appearing in the
expressions given above. Zero-field perturbation-strength
differentiation of the orthonormality, TDSCF, and \ac{FQ} equations yields the
desired response equations.\cite{Thorvaldsen2008-sg,Ringholm2014-gx,diremigio2017open}
Solution of the $N$-th order response equations for the $b_{\mathrm{N}}$
perturbation tuple yields the desired response parameters $\rspParam{b_{N}}$.
The perturbed variational parameters are further partitioned into a sum of \emph{homogeneous} and
\emph{particular} contributions.\cite{Thorvaldsen2008-sg} Whereas the former depend on the $N$-th order
response parameters, the latter depend only on lower order response parameters.
With this partition, the response equations to any order can be compactly rearranged as:\cite{Larsen2000-hj, Kjaergaard2008-hy}
\begin{equation}
\left[ \genHessian - \omega_{b_N} \genMetric \right] \rspParam{b_{N}} = \RHS{b_{N}}.
\end{equation}
The \ac{LHS} includes the generalized Hessian and metric matrices, $\genHessian$
and $\genMetric$, respectively and $\omega_{b_N} = \sum_{i=1}^{N}\omega_{b_i}$.
The \ac{RHS} $\RHS{b_{N}}$ collects contributions from lower-order perturbed
density matrices and $N$-th order particular contributions, see Refs.~\citenum{Thorvaldsen2008-sg,Ringholm2014-gx,diremigio2017open}.

For a single, electric-field type perturbation, the matrix-vector products
$\genHessian\rspParam{b}$ and $\genMetric\rspParam{b}$ assume the form:
\begin{align}
  \genHessian\rspParam{b} &=
  \mat{G}^\mathrm{KS}([\rspParam{b}, \mat{D}]_{\mat{S}})\mat{D}\mat{S}
  -\mat{S}\mat{D}\mat{G}^\mathrm{KS}([\rspParam{b}, \mat{D}]_{\mat{S}})
+\mat{F} [\rspParam{b}, \mat{D}]_{\mat{S}} \mat{S}\nonumber \\
&-\mat{S} [\rspParam{b}, \mat{D}]_{\mat{S}} \mat{F}
+ \mat{q}_{\mathrm{H}}^b\mat{V}\mat{D}\mat{S}
-\mat{S}\mat{D}\mat{V}\mat{q}^b_{\mathrm{H}}
\label{eq:E2-lintra}
\\
  \genMetric\rspParam{b} &= \mat{S} [\rspParam{b}, \mat{D}]_{\mat{S}} \mat{S} \text{.}
  \label{eq:S2-lintra}
\end{align}
where $\mat{G}^\mathrm{KS}$ now collects the two-electron and exchange-correlation contributions.
The general expression for the \ac{RHS} (see eq.~(46) in Ref.~\citenum{diremigio2017open})
simplifies to the matrix elements of the
electric dipole perturbation operator, with no contributions from the classical
polarizable model.
These equations are equivalent, upon transformation to the \ac{MO} basis, to their
more familiar formulation as Casida's equations.\cite{Stratmann1998-ve,lipparini2012linear}
Perturbed \acp{FQ} are obtained by solving:
\begin{equation}
  \mat{J}\mat{q}^{b}_{\mathrm{H}}
=
 - \mat{V}(\mat{D}^{b}_{\mathrm{H}}),
\end{equation}
once again highlighting the introduction of the mutual QM/MM polarization.

Response equations for the second-order response parameters $\rspParam{bc}$ can
be derived in a similar fashion. Restricting ourselves to electric-field type
perturbations only, the linear transformations are expressed as:
\begin{align}
  \genHessian\rspParam{bc} &=
  \mat{G}^\mathrm{KS}([\rspParam{bc}, \mat{D}]_{\mat{S}})\mat{D}\mat{S}
  -\mat{S}\mat{D}\mat{G}^\mathrm{KS}([\rspParam{bc}, \mat{D}]_{\mat{S}})
+\mat{F} [\rspParam{bc}, \mat{D}]_{\mat{S}} \mat{S}\nonumber \\
&-\mat{S} [\rspParam{bc}, \mat{D}]_{\mat{S}} \mat{F}
+ \mat{q}_{\mathrm{H}}^{bc}\mat{V}\mat{D}\mat{S}
-\mat{S}\mat{D}\mat{V}\mat{q}^{bc}_{\mathrm{H}}
\\
  \genMetric\rspParam{bc} &= \mat{S} [\rspParam{bc}, \mat{D}]_{\mat{S}} \mat{S} \text{.}
\end{align}
In contrast to the linear response equations, the \ac{RHS} will contain
\ac{FQ} contributions:
\begin{equation}
  \mat{M}_\mathrm{RHS,FQ}^{bc} = \aherm{
    (\mat{q}^{bc}_{\mathrm{P}})^{\dagger}\mat{V}\mat{D}\mat{S}
  + \mat{q}^{\dagger}\mat{V}\Dp{bc}\mat{S}
  + (\mat{q}^{b}_{\mathrm{H}})^{\dagger}\mat{V}\mat{D}^{c}_{\omega}\mat{S}
  + (\mat{q}^{c}_{\mathrm{H}})^{\dagger}\mat{V}\mat{D}^{b}_{\omega}\mat{S}
  },
\end{equation}
with the second-order particular \acp{FQ} calculated as the solution to the
linear equation:
\begin{equation}
  \mat{J}\mat{q}^{bc}_{\mathrm{P}}
=
  - \mat{V}(\mat{D}^{bc}_{\mathrm{P}}),
\end{equation}
and the perturbed density matrix $\mat{D}^{bc}_{\mathrm{P}}$ is in turn assembled
from first-order perturbed density matrices.

As for the \ac{PCM}, there are two classes of contributions from the classical
polarizable region: implicit, through the unperturbed Fock matrix, and explicit,
through the $N$-th order perturbed homogeneous \acp{FQ}.
This is indeed a trait shared by any quantum/classical polarizable model.

\subsubsection{One- and Two-Photon Absorption}

We can formulate one- and two-photon absorption parameters in terms of single residues
of the linear and quadratic response functions, respectively.
\citet{Friese2015-kb} have presented a density matrix-based, open-ended
formulation of single residues that can be coupled to classical polarizable
models.\cite{steindal2016open,diremigio2017open}

TPA cross sections of randomly oriented systems can be calculated from the
imaginary part of the third susceptibility. Alternatively, they can be
obtained as the individual two-photon transition matrix elements $S_{ab}$
between the initial state $\ket{i}$ and final state $\ket{f}$, with the
sum-over-states expression:\cite{cronstrand2005multi}
\begin{equation}\label{eq:trans_mom_tpa}
S_{ab} = \sum_s \left( \frac{\braket{i|\mu_a|s}\braket{s|\mu_b|f}}{\omega_{si}-\omega} + \frac{\braket{i|\mu_b|s}\braket{s|\mu_a|f}}{\omega_{si}-\omega}\right)
\end{equation}
where $a,b \in x,y,z$ and $\omega$ is the frequency of the external radiation,
which is half of the excitation energy $\omega_f$ to the final state $\ket{f}$
($2\omega = \omega_f$). The summation runs over all $s$ states, including
initial and final state. $\omega_{si} = \omega_s - \omega_i$ is the transition
energy between $s$ and $i$ states.

For linearly polarized light with parallel polarization, rotationally averaged
microscopic \ac{OPA} and \ac{TPA} cross sections can be written in terms of the
transition matrix tensors $\mathbf{S}$ and their complex conjugates
$\mathbf{\overline{S}}$ as:
\begin{align}
\langle \delta^{\rm 1PA} \rangle &= \frac{1}{3}\sum_{a} S_{a} \bar{S}_{a} \\
\langle \delta^{\rm 2PA} \rangle &= \frac{1}{15}\sum_{ab} \left( 2 S_{ab} \bar{S}_{ab} + S_{aa} \bar{S}_{bb} \right).
\end{align}

Finally, the macroscopic \ac{TPA} cross section in cgs units can be obtained from the
rotationally averaged \ac{TPA} strengths ($\braket{\delta^{\text{TPA}}}$) expressed
in atomic units as:
\begin{equation}
\sigma^{\text{TPA}} = \frac{N\pi^3\alpha a^5_0\omega^2}{c}\braket{\delta^{\text{TPA}}} g(2\omega,\omega_0,\Gamma)
\end{equation}
where $N=4$ in case of single beam experiments, $\alpha$ is the fine structure constant, $a_0$ is
the Bohr radius, $\omega$ is the photon energy in atomic units, $c$ is the speed of light and
$g(2\omega,\omega_0,\Gamma)$ the lineshape function describing spectral broadening effects.
The common unit for \ac{TPA} cross sections is GM in honour of the work of Maria Goppert-Mayer
(1 GM = 10$^{-50}$ cm$^4$ s photon$^{-1}$). We refer the reader to Ref. \citenum{beerepoot2015benchmarking} for further details on the computational approach to TPA cross sections.

\subsection{Implementation}

The close similarity of the \ac{PCM} and \ac{FQ} models renders the
implementation of the latter a rather straightforward extension of the former.
Our implementation relies on the PCMSolver library,\cite{diremigio2019pcmsolver}
which already provides the infrastructure for the \ac{PCM} and lends itself to
the extensions needed for a \ac{FQ} implementation.
The calculation of TPA cross sections with the \ac{FQ} discrete model was
implemented in the LSDALTON program package.\cite{Aidas2014-rp}
LSDALTON provides single residues of the quadratic response function \emph{in
vacuo} and we exploited the existing interface with
PCMSolver\cite{bugeanu2015wavelet} to extend this functionality to include the
\ac{PCM} and \ac{FQ} quantum/classical polarizable models.
The TPA implementation in LSDALTON leverages the $2n+1$ rule. Ensuring
correctness of the linear response calculations is enough to guarantee
correctness up to cubic response properties.
We performed extensive testing of the \ac{PCM} and \ac{FQ} linear and quadratic response
functions, including their residues, against DALTON.\cite{frediani2005two,zhao2007solvent,frediani2005second,ferrighi2007two}

The use of a modular programming paradigm, based on open-source libraries and
programs, is particularly beneficial for the work here presented. The most
significant programming investment was the extension of the PCMSolver library.

\section{Computational Details}\label{sec:comp-details}

\begin{figure}
\centering
\includegraphics[width=.5\textwidth]{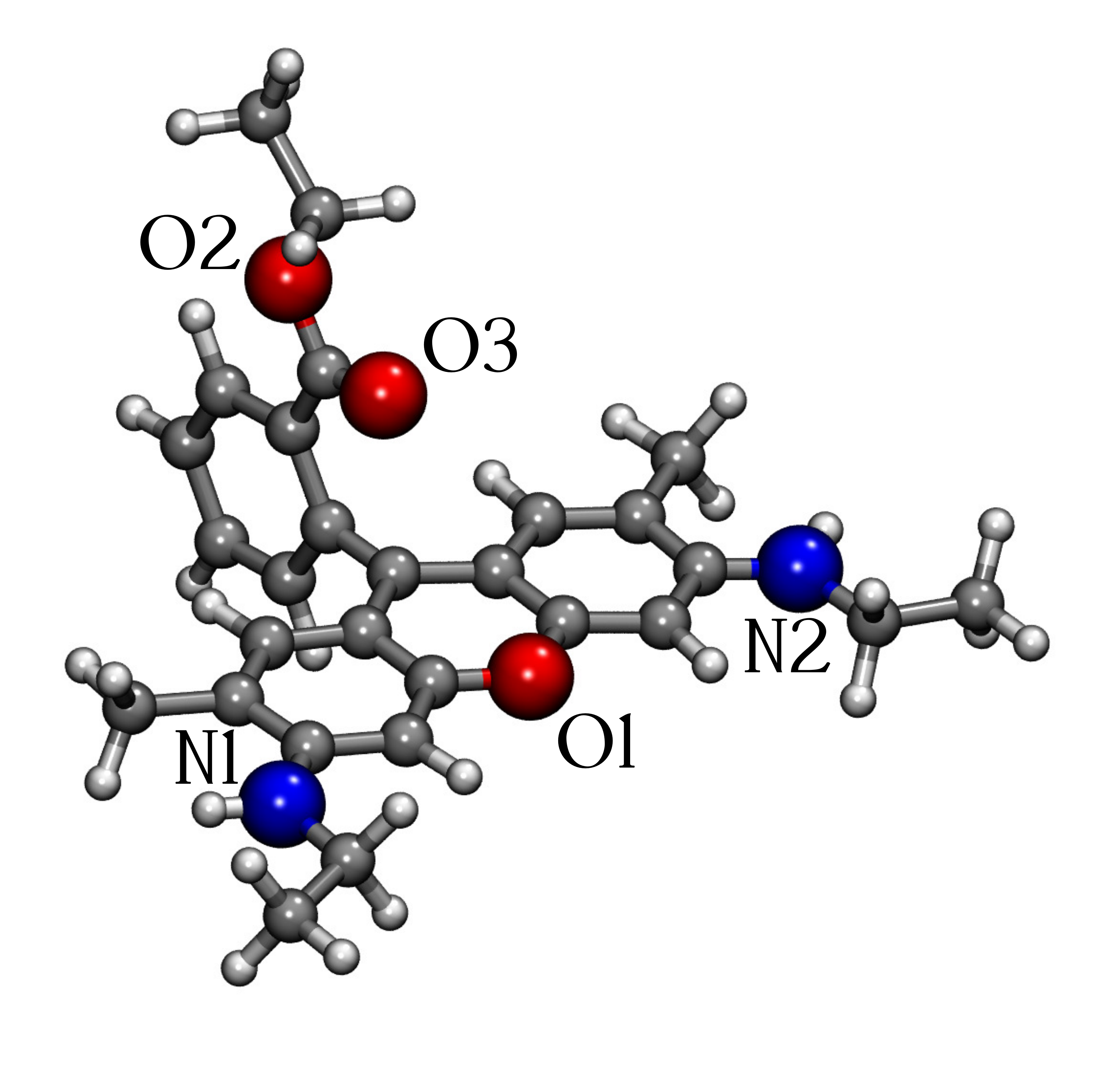}
\caption{Rhodamine 6G structure and atom labeling.}
\label{fig:structureR6G}
\end{figure}

Rhodamine 6G (R6G) (see Figure \ref{fig:structureR6G}) in aqueous solution
has been amply studied experimentally using both OPA and TPA
techniques.\cite{makarov2008two,milojevich2013surface,milojevich2011probing,guthmuller2008resonance1,
guthmuller2008resonance2,dieringer2008surface,weiss2014non,nag2009solvent}
Such a system is capable of forming solute-solvent hydrogen bonds, and is
thus a good test case for our atomistic polarizable QM/FQ approach.

We adopted the following computational protocol for our QM/MM calculations of
excitation energies, OPA and TPA intensities:
\begin{description}
\item[Definition of the systems and calculation of atomic charges] The solute
molecule was surrounded by a number of water molecules large enough to
represent all the solute-solvent interactions (at least 8500). The atomic charges of the solute
were computed by using the Charge Model 5 (CM5).\cite{dodda2015evaluation}
\item[Classical MD simulations in aqueous solution] The MD simulations were
performed in a cubic box reproduced periodically in every direction, satisfying
periodic boundary conditions (PBC). A minimization step ensures that
simulations were started from a minimum of the classical PES. From the
MD run, a set of snapshots was extracted to be used in the following
QM/TIP3P\cite{mark2001structure} and QM/FQ calculations.
\item[Definition of the different regions of the two-layer scheme and their
boundaries] Each snapshot extracted from the MD run was cut into a sphere
centered on the solute. A radius of 25 \AA~ was chosen in order to include
all specific water-solute interactions.
\item[QM/classical calculations] QM/TIP3P, and QM/FQ OPA
and TPA calculations were performed on 100 structures obtained in the
previous step. The results obtained for each spherical snapshot were extracted
and averaged to produce the final value.
\end{description}

In step 1, \ac{R6G} was optimized and CM5 charges were calculated at the
B3LYP/6-31+G* level of theory including solvent effects by means of the PCM
\cite{tomasi2005}.

The MD simulation was performed using GROMACS\cite{Gromacs5}, with the
OPLS-AA \cite{gromos} force fields to describe intra- and inter-molecular
interactions. CM5 charges were used to account for electrostatic interactions.
The TIP3P FF was used to describe the water molecules \cite{mark2001structure}.
A single molecule was dissolved in a cubic box containing at least 8500 water
molecules. A chloride ion has been included in the box to neutralize the system.
The chromophore was kept fixed during all the steps of the MD run. This neglects
geometric relaxation due to solvation and is an approximation. However, our
primary goal is to compare \emph{direct} (electronic) rather than
\emph{indirect} (geometrical) solvent effects on the properties. Furthermore,
this approximation affords a fairer comparison with implicit solvent models.
\ac{R6G} in aqueous solution was initially brought to 0 K with the steepest
descent minimization procedure and then heated to 298.15 K in an NVT ensemble
using the velocity-rescaling\cite{vrescale} method with an integration time
step of 0.2 fs and a coupling constant of 0.1 ps for 200 ps. The time step and
temperature coupling constant were then increased to 2.0 fs and 0.2 ps,
respectively, and an NPT simulation (using the Berendsen barostat and a coupling
constant of 1.0 ps) for 1 ns was performed to obtain a uniform distribution of
molecules in the box. A 10 ns production run in the NVT ensemble was then
carried out, fixing the fastest internal degrees of freedom by means of the
LINCS algorithm ($\delta$t=2.0 fs)\cite{hess1997lincs}, and freezing the
\ac{R6G} at the center of the simulation box. Electrostatic interactions are
treated by using particle-mesh Ewald (PME) \cite{darden1993particle} method with
a grid spacing of 1.2 \AA~and a spline interpolation of order 4. We have
excluded intramolecular interactions between atom pairs separated up to three
bonds. A snapshot every 100 ps was extracted in order to obtain a total of 100
uncorrelated snapshots.

For each snapshot a solute-centered sphere with a radius of 25 \AA~was cut.
Notice that the chloride ion was not present in any of the
extracted spherical snapshots. For each snapshot, OPA and TPA spectra were then
calculated with two QM/MM approaches: the water molecules were modeled by means of the
non-polarizable TIP3P FF,\cite{mark2001structure} and the FQ SPC parametrization
proposed by \citet{rick1994dynamical}.
For comparison, we also ran QM calculations on the isolated chromophore and
embedded in a \ac{PCM} continuum modelling the water solution.

We performed all \ac{OPA} calculations using a locally modified version of the
Gaussian 16 package,\cite{gaussian16} whereas we used a locally modified version
of the LSDALTON program,\cite{Aidas2014-rp} interfaced to the PCMSolver
library,\cite{diremigio2019pcmsolver} for the \ac{TPA} calculations.
The CAM-B3LYP/6-31+G* model chemistry was used in all calculations. For the
LSDALTON calculations we leveraged the implementation of density fitting, with
the df-def2 auxiliary fitting basis, to accelerate the evaluation of the Coulomb
matrix.

For the \ac{PCM} calculations, the cavity was generated from a set of
atom-centered, interlocking spheres.
PCMSolver implements the GePol algorithm for cavity generation and uses the
Bondi--Mantina set of van der Waals radii~\cite{Bondi1964-dt, Mantina2009-hb} 1.20~\AA{} for hydrogen, 1.70~\AA{} for carbon,
1.55~\AA{} for nitrogen and 1.52~\AA{} for oxygen. All radii were scaled by a factor of 1.2.
Values of the static and optical permittivities of $\diels=78.39$ and
$\dield=1.776$, respectively, were used for the \ac{PCM} calculations in water
in LSDALTON.

For the \ac{PCM} calculations in Gaussian, the following atomic radii were used and scaled by a factor of 1.1:
1.443~\AA{} for hydrogen, 1.9255~\AA{} for carbon, 1.83~\AA{} for nitrogen and 1.75~\AA{} for oxygen.
Values of the static and optical permittivities of $\diels=78.3553$ and
$\dield=1.77785$, respectively, were instead used to model water solvent effects.

The differences in \ac{PCM} parametrization between the two codes used in this
study did not lead to significant differences (less than 1\%) between computed
ground state energies, excitation frequencies and oscillator strengths.

\section{Numerical Results}\label{sec:results}

First, we analyze the MD runs in terms of
the hydration patterns with a particular focus on hydrogen bonds (HBs) formed
between \ac{R6G} and solvent water molecules. Second, OPA and TPA spectra are
presented and compared with their experimental counterparts.

\subsection{MD analysis: Hydration Pattern}

\begin{figure}
\centering
\includegraphics[width=.8\textwidth]{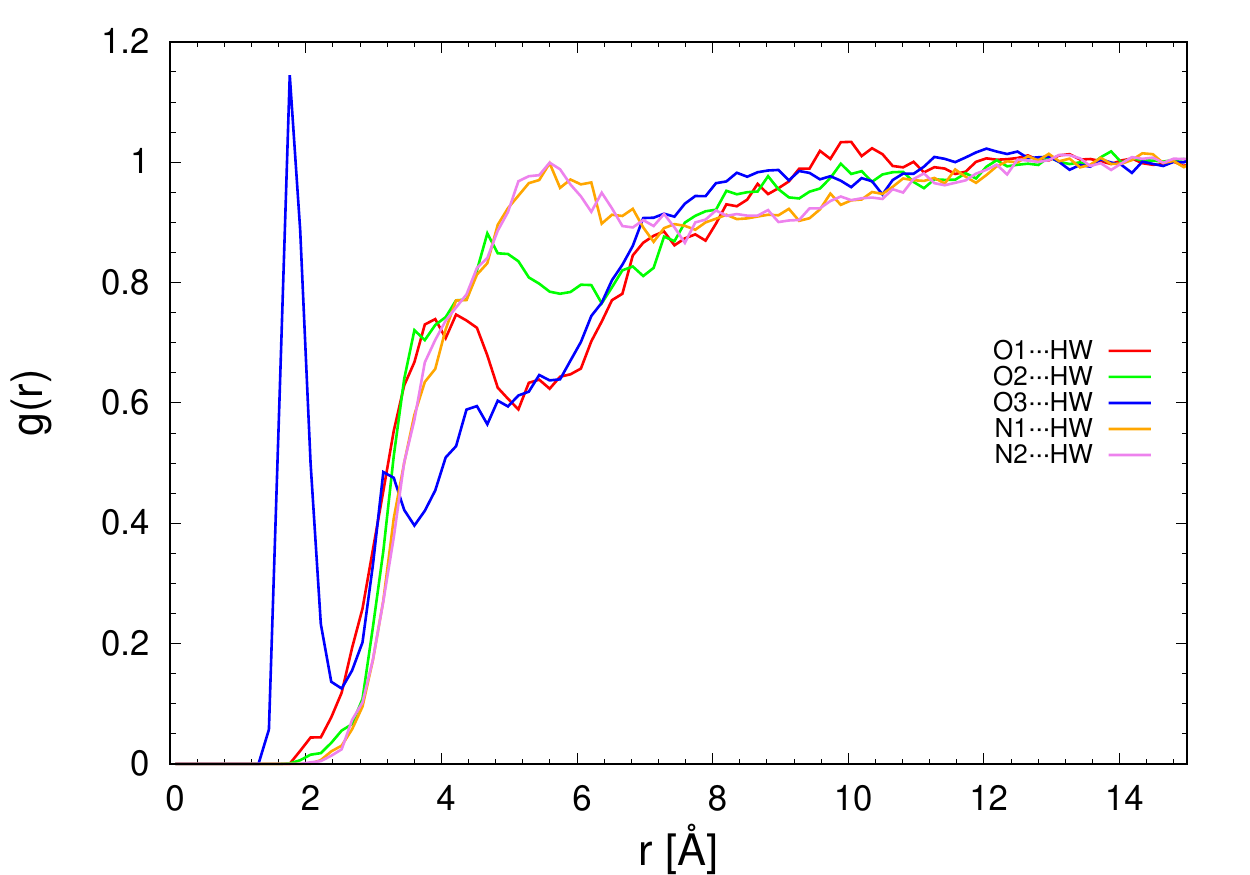}
\caption{Radial distribution function $g(r)$ between selected sites of \ac{R6G} and water molecules. The considered atomic sites are highlighted in Figure \ref{fig:structureR6G}}
\label{fig:gdr}
\end{figure}

\ac{R6G} is characterized by a keto oxygen and by amino groups, which can act
as HB donors, whereas the ether oxygen atoms (O1 and O2) together with the amino
nitrogen atoms (N1 and N2) can act as HB acceptors (see Figure \ref{fig:structureR6G}
for atom labeling).
HB patterns were analyzed by extracting the radial distribution functions $g(r)$
from the MD trajectories. For this analysis the TRAVIS
package was used.\cite{brehm2011travis}
The radial distribution functions were computed taking as reference oxygen atoms (O1,
O2 and O3), nitrogen atoms (N1 and N2) and amino hydrogen atoms (H1 and H2) of the
solute; they are plotted in Figure \ref{fig:gdr}.
The most intense peak of the $g(r)$ refers to the carboxylic oxygen (O3),
whereas the other atoms are not involved in HBs with the solvent water
molecules. The average number of HBs between water molecules and the carboxylic
oxygen (O3) is 1.7, thus confirming strong HB patterns.

\subsection{OPA Spectra of \ac{R6G} in aqueous solution}

100 uncorrelated snapshots were extracted from the MD run. The convergence of
the studied properties (OPA and TPA spectra) was checked
by considering an increasing number of snapshots. These results are reported in
Figures S1 and S2 in the \ac{SI}.

\begin{figure}[htbp!]
\centering
\includegraphics[width=.8\textwidth]{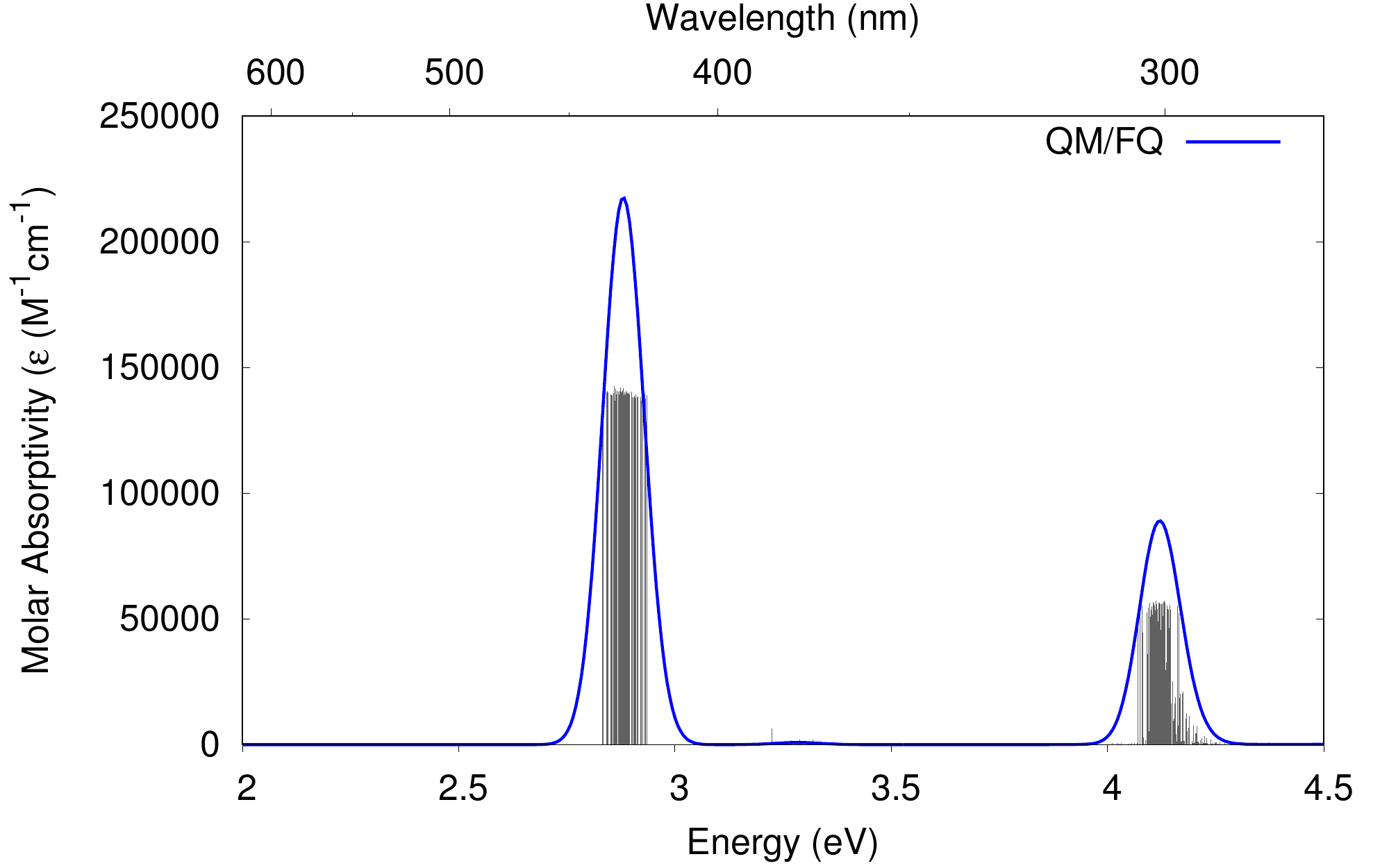}
\caption{QM/FQ calculated OPA spectrum of \ac{R6G} in aqueous solution reported as stick spectrum and convoluted with a Gaussian band shape (FWHM=0.1 eV)}
\label{fig:FQ-stick-OPA}
\end{figure}

\ac{QM}/\ac{FQ} \ac{OPA} results are depicted in Figure \ref{fig:FQ-stick-OPA}.
Raw data (sticks) and their Gaussian convolution, with full width at half maximum
(FWHM) of 0.1 eV, are shown.

\begin{figure}
\centering
\includegraphics[width=.5\textwidth]{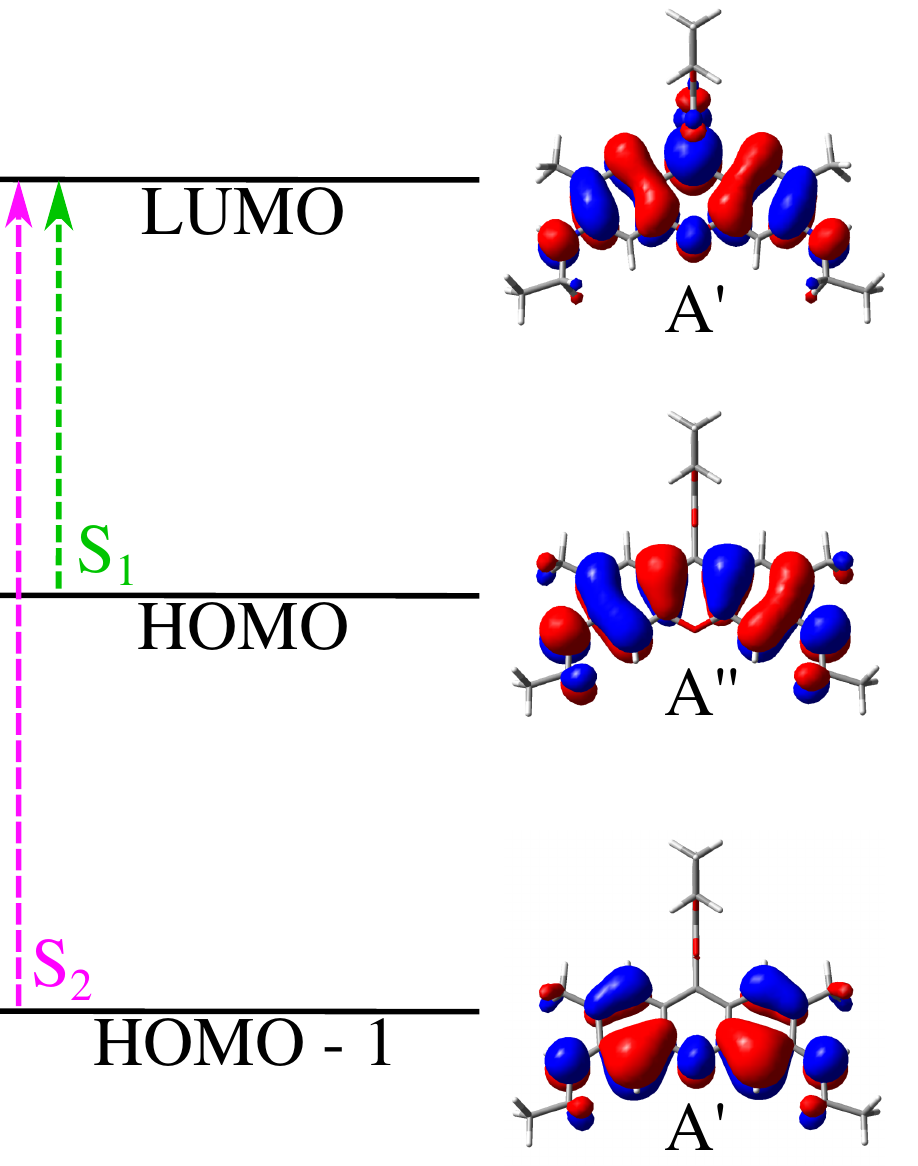}
\caption{Graphical scheme of the first two transitions of \ac{R6G} in aqueous solution. Molecular orbitals related to each transition are depicted.}
\label{fig:orbitals}
\end{figure}

The OPA spectrum is characterized by an intense transition (S1) at about 2.8 eV
(440 nm) which is related to a pure HOMO$\rightarrow$LUMO transition. The second
transition (S2) is a pure HOMO-1$\rightarrow$LUMO transition and is located at
about 3.2 eV (380 nm). This transition is dark due to the symmetry of the
involved orbitals, see Figure \ref{fig:orbitals}.
Our findings confirm what already
reported in previous theoretical studies,\cite{milojevich2013surface} and is in
contrast with some experimental works,\cite{nag2009solvent} where S2 is instead assigned to a visible transition.
The third transition predicted by the QM/FQ approach is located at
4.1 eV (300 nm), and involves a combined transition between
HOMO-1$\rightarrow$LUMO+1 and HOMO-2$\rightarrow$LUMO. The whole computed
spectrum, involving also the higher transitions, is reported in the SI (
Figure S3).

\begin{figure}[htbp!]
\centering
\includegraphics[width=.8\textwidth]{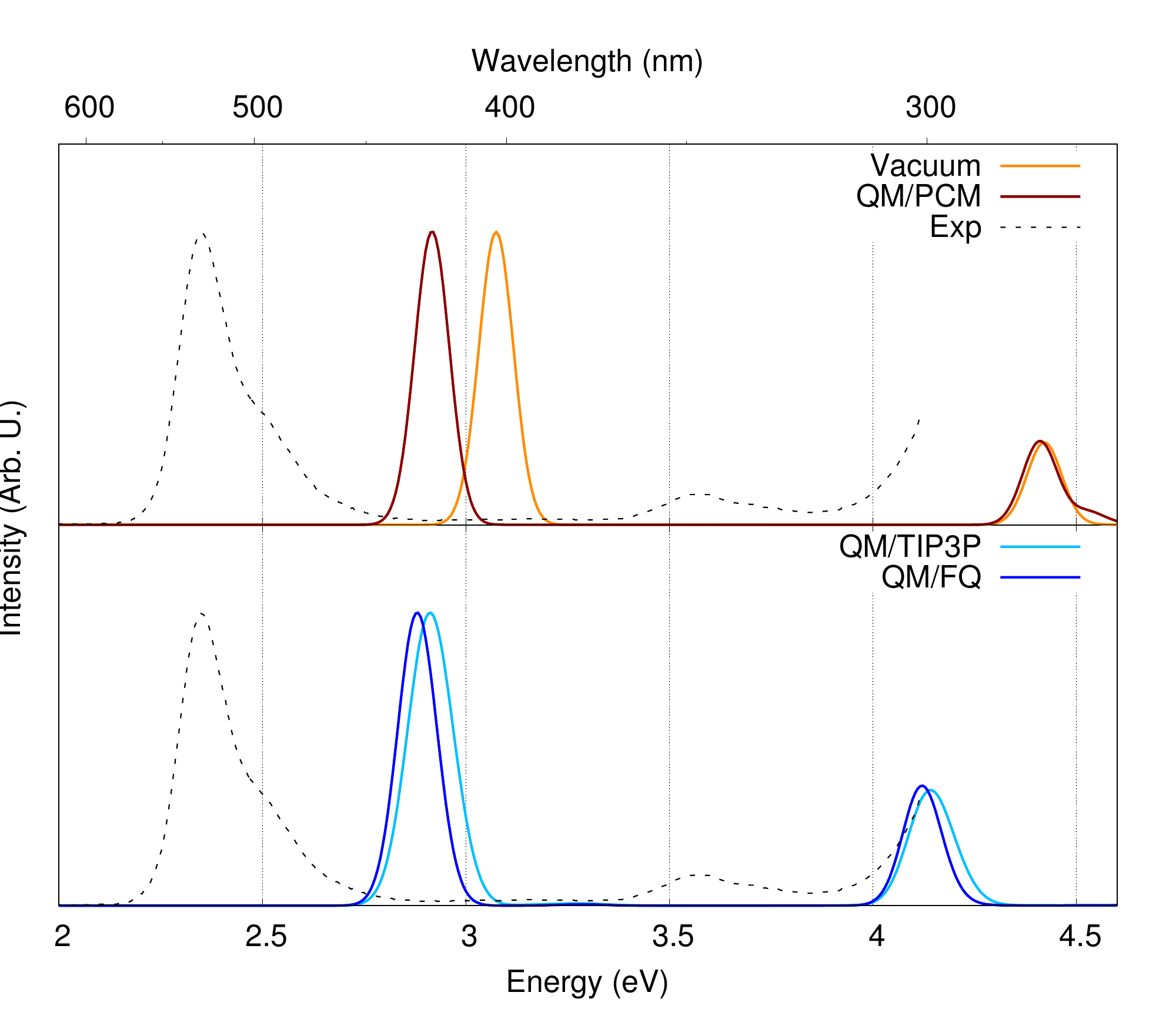}
\caption{One-photon absorption spectra for the \ac{R6G} chromophore at the
CAM-B3LYP/6-31+G* level of theory. Computed vacuum, QM/PCM, QM/TIP3P and QM/FQ
OPA spectra are reported. The experimental spectrum is reproduced from
Ref.~\citenum{nag2009solvent}.}
\label{fig:OPA-methods}
\end{figure}

We investigated the relevance of solvent effects by employing different
computational approaches, ranging from continuum QM/PCM to nonpolarizable QM/TIP3P and
polarizable QM/FQ descriptions.
As a reference, the \ac{OPA} spectrum of the solute \emph{in vacuo} was also
calculated.
Vertical excitations and oscillator strengths as obtained by exploiting the
different approaches summarized in Table \ref{tab:OPA-methods} and
graphically depicted in Figure \ref{fig:OPA-methods}. The experimental spectrum
is also reported.\cite{nag2009solvent}

\begin{table}[htbp!]
\centering
\caption{Vacuum and aqueous solution vertical excitation energies (eV) and oscillator
strengths (a.u.) for the three lowest singlet excited states of \ac{R6G} calculated at
the CAM-B3LYP/6-31+G* level of theory. QM/PCM, QM/TIP3P and QM/FQ approaches
were used to model environment effects in solution. The experimental vertical
excitation energies are also reported.}
\label{tab:OPA-methods}
\begin{tabular}{c S S S S S S S S S}
\toprule
  & \multicolumn{2}{c}{Vacuum} & \multicolumn{2}{c}{QM/PCM} & \multicolumn{2}{c}{QM/TIP3P} & \multicolumn{2}{c}{QM/FQ} & {Exp.\text{$\phantom{t}^{\text{a}}$}} \\
\cmidrule(lr){2-10}
  & {$E_{\text{vert}}$} & {$f$} & {$E_{\text{vert}}$} & {$f$} & {$E_{\text{vert}}$} & {$f$} & {$E_{\text{vert}}$} & {$f$} & {$E_{\text{vert}}$} \\
\cmidrule(lr){1-10}
S1 & 3.07 & 0.89 & 2.98  &  1.05 &  2.82 &  0.84 & 2.81  &  0.90 &  2.27 \\
S2 & 3.68 & 0.00 & 3.71  &  0.00 &  3.18 &  0.01 & 3.20  &  0.00 &       \\
S3 & 4.42 & 0.25 & 4.47  &  0.29 &  4.15 &  0.21 & 4.13  &  0.26 &  3.46 \\
\bottomrule
\end{tabular}

\text{$\phantom{t}^{\text{a}}$} Experimental values are taken from Ref. \citenum{nag2009solvent}.\hfill
\end{table}

The experimental spectrum is dominated by an intense transition (S1) at about
2.3 eV (550 nm), characterized by an inhomogeneous band broadening,
probably due to a vibronic convolution. The second visible transition
is instead at about 3.6 eV (350 nm) and can be associated to the S3 transition.

The first transition (S1) is predicted to be the most intense by all the
different methods, with a redshift when solvent effects are taken into
consideration. In particular, QM/PCM, QM/TIP3P and QM/FQ predict very similar
vertical excitation energies for the S1 transition, with the largest redshift
shown by QM/FQ.
All the approaches considered correctly describe the S2 transition as being
symmetry-forbidden, see Table \ref{tab:OPA-methods}.
The models employed exhibit major differences for the third transition (S3), which
is predicted at about 280 nm in vacuum and by QM/PCM, and at about 310 nm by
both QM/TIP3P and QM/FQ approaches (see Fig.\ref{fig:OPA-methods}).
The atomistic description captures the explicit solvent-solute interactions 
that appear to be crucial in the modeling of this transition.
Implicit solvation is unable to capture solvent effects, as it is shown by
the negligible differences observed with respect to the vacuum results.
Furthermore, the energy difference between S1 and S3 is
correctly predicted by QM/MM approaches ($\sim 1.18$ eV) as compared to the
experimental value (1.19 eV), whereas it is overestimated by the QM/PCM approach
(1.49 eV). Notice also that by considering energy differences, instead of absolute
vertical excitation energies, systematic errors and biases due to the choice of
QM method, and in particular of a specific DFT functional, should be avoided.

To further analyze the nature of the electronic transitions, their charge
transfer (CT) nature was characterized by a simple index, denoted as
$D_{\mathrm{CT}}$.\cite{le2011qualitative,egidi2018nature} The barycenters of the
positive and negative density distributions are calculated as the difference of
ground state (GS) and excited state (ES) densities. The CT length index
($D_{\mathrm{CT}}$ ) is defined as the distance between the two barycenters. Calculated
vacuum, QM/PCM, QM/TIP3P and QM/FQ $D_{\mathrm{CT}}$ values are reported in Table
\ref{tab:dct-index}.

\begin{table}[htbp!]
\centering
\caption{$D_{\mathrm{CT}}$ indices (\AA) for the first three excited states as obtained by exploiting the various models considered in this work.}
\label{tab:dct-index}
\begin{tabular}{c S S S S}
\toprule
       & {Vacuum} & {QM/PCM} & {QM/TIP3P} & {QM/FQ} \\
\cmidrule(lr){2-5}
S1 & 1.195  & 1.170  &  1.668 & 1.624  \\
S2 & 1.318  & 1.336  &  1.587 & 1.574  \\
S3 & 0.811  & 0.759  &  4.043 & 3.580  \\
\bottomrule
\end{tabular}
\end{table}

The first two transitions (S1 and S2) have little CT character. The largest difference between the considered
approaches is shown by the S3 transition, of which the CT character is irrelevant \emph{in vacuo} and at the QM/PCM level, whereas is huge at both QM/TIP3P and QM/FQ levels. This different behaviour was expected
by considering the computed OPA spectra reported in Figure \ref{fig:OPA-methods}; in fact, the largest discrepancy between vacumm-QM/PCM and QM/MM
approaches was indeed predicted for the S3 transition.

\subsection{TPA Spectra of \ac{R6G} in aqueous solution}

\begin{figure}[htbp!]
\centering
\includegraphics[width=.8\textwidth]{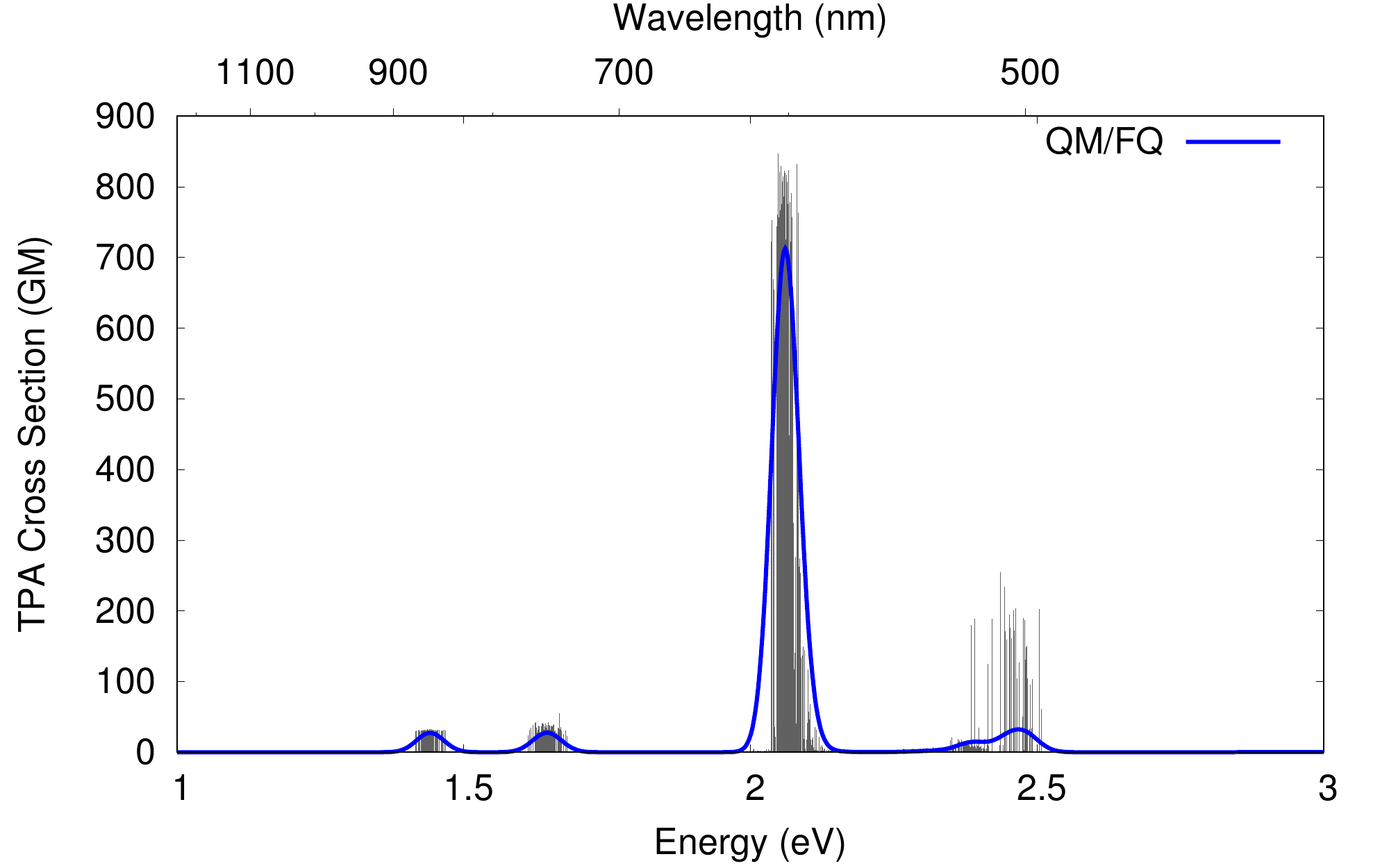}
\caption{QM/FQ calculated TPA spectrum of \ac{R6G} in aqueous solution reported
as stick spectrum and convoluted with a Gaussian band shape (FWHM = 0.1 eV)}
\label{fig:FQ-stick-TPA}
\end{figure}

Figure \ref{fig:FQ-stick-TPA} reports computed QM/FQ TPA raw
data and their Gaussian convolution. We
adopted a Gaussian lineshape with FWHM of 0.1 eV in agreement with previous
computational studies on TPA and suggested best practices, see Ref.~\citenum{beerepoot2015benchmarking}.

The first two visible transitions at about 1.4 and 1.6 eV have almost the same
intensity and they are the S1 and S2 transitions, respectively. The S2 transition is not
dark, due to the different symmetry selection rules in TPA compared to OPA (see
Figure \ref{fig:FQ-stick-OPA}).
The computed QM/FQ TPA spectrum is dominated by an intense peak at about 2.06 eV
(600 nm) associated to the S3 transition.

\begin{figure}[htbp!]
\centering
\includegraphics[width=.8\textwidth]{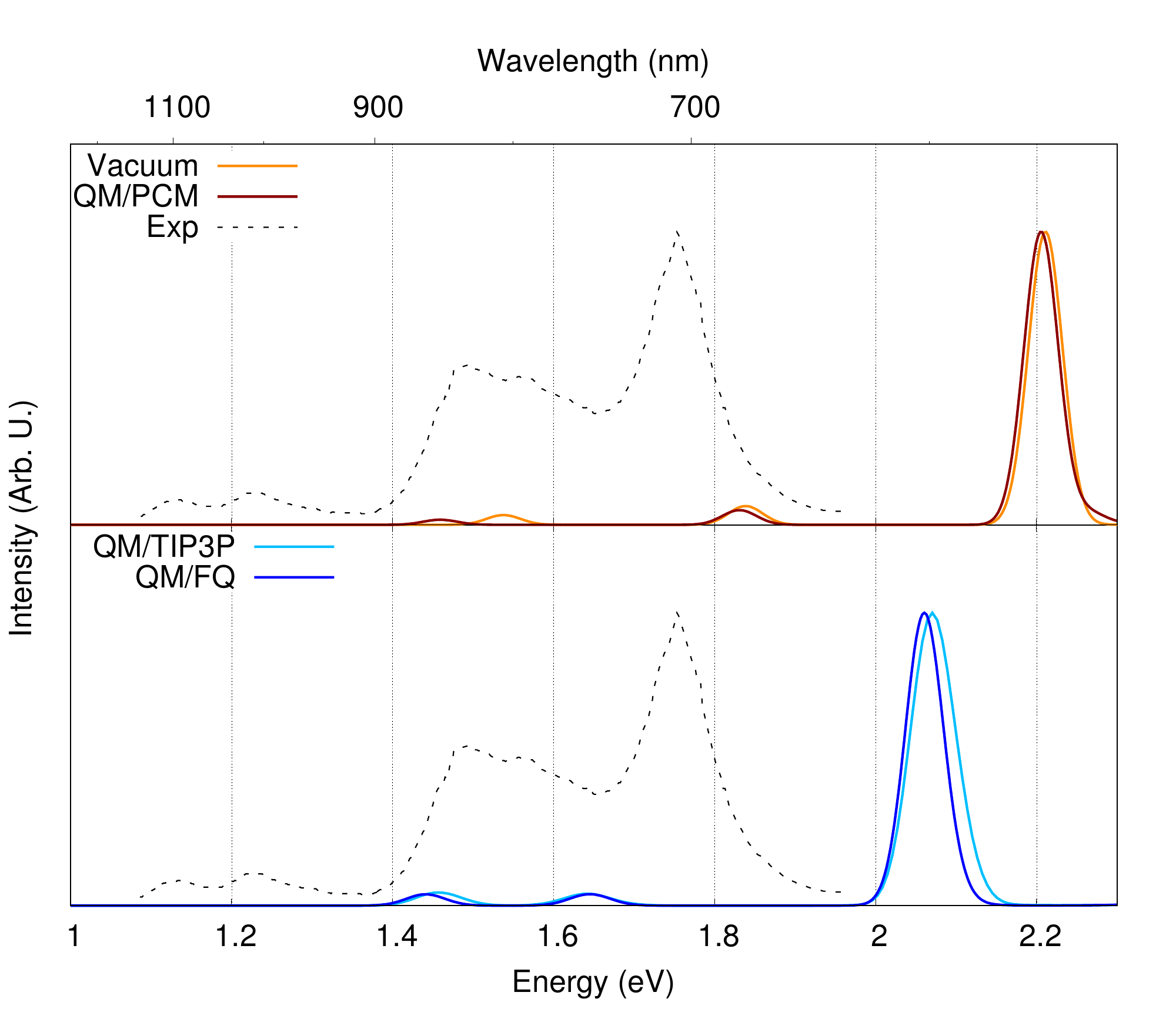}
\caption{Two-photon absorption spectra for the R6G chromophore at the CAM-
B3LYP/6-31+G* level of theory. Computed vacuum, QM/PCM, QM/TIP3P and QM/FQ
OPA spectra are reported. The experimental spectrum is reproduced from
Ref.~\citenum{makarov2008two}.}
\label{fig:TPA-methods}
\end{figure}

We compared different approaches to model solvent effects also for TPA
spectra and considered QM/PCM, QM/TIP3P and QM/FQ models.
In addition, TPA spectra of \ac{R6G} \emph{in vacuo} was computed as an
additional reference point.
Vacuum, QM/PCM, QM/TIP3P and QM/FQ TPA spectra are plotted in Figure
\ref{fig:TPA-methods} together with the experimental spectrum, reproduced from
Ref.~\citenum{makarov2008two}.

The experimental TPA spectrum is characterized by three main peaks at 1.14,
1.52, and 1.73 eV. These correspond to the S1, S2 and S3 transitions,
respectively.
It is worth noting that the peak at about 1.25 eV was wrongly reported to be
associated to the S2 transition in some experimental works due to the similar
intensity with respect to the peak at 1.14 eV.\cite{nag2009solvent}
\citeauthor{milojevich2013surface} recently showed that such a peak is instead a
vibronic band due to Herzberg--Teller terms.\cite{milojevich2013surface}
Notice in fact that such a vibronic peak exactly corresponds to the vibronic peak
present also in the experimental OPA spectrum, see Figure \ref{fig:OPA-methods}.
In this work, we are not considering any vibronic contributions because our main
goal is to show the performance of QM/FQ approach in predicting solvent effects
on multiphoton spectroscopies. As a consequence, the peak at 1.25 eV cannot be
reproduced by the different approaches explored in this work.

The computed vacuum and aqueous solution transition energies and TPA cross sections
(in GM units) of the first three transitions are reported in Table
\ref{tab:TPA-methods}, together with their experimental values.
It is clear from Figure \ref{fig:TPA-methods} and Table \ref{tab:TPA-methods}
that all the approaches considered in this work predict the S3 transition as the
most intense in the TPA spectrum. The whole TPA spectra, including also higher energy
transitions, are reported in Figure S4 of the SI.

\begin{table}[htbp!]
\centering
\caption{Vacuum and aqueous solution TPA vertical excitation energies (eV) and
cross sections (GM) for the six lowest singlet excited states of \ac{R6G}
calculated at the CAM-B3LYP/6-31+G* level of theory. QM/PCM, QM/TIP3P and
QM/FQ approaches were used to model environment effects in
aqueous solution. The experimental TPA vertical excitation energies and cross sections
are also reported.}
\label{tab:TPA-methods}
\begin{tabular}{c S S S S S S S S S S}
\toprule
       & \multicolumn{2}{c}{Vacuum} & \multicolumn{2}{c}{QM/PCM} & \multicolumn{2}{c}{QM/TIP3P} & \multicolumn{2}{c}{QM/FQ} & \multicolumn{2}{c}{Exp.\text{$\phantom{t}^{\text{a}}$}} \\
\cmidrule(lr){2-11}
       & {$E_{\text{TPA}}$} & {$\sigma$} & {$E_{\text{TPA}}$} & {$\sigma$} & {$E_{\text{TPA}}$} & {$\sigma$} & {$E_{\text{TPA}}$} & {$\sigma$} & {$E_{\text{TPA}}$} & {$\sigma$} \\
\cmidrule(lr){1-11}
S1 & 1.54 &  15.52  & 1.46 &   24.03 & 1.46 &  18.67 & 1.44 &  27.37 & 1.14 &  15 \\
S2 & 1.84 &  29.83  & 1.83 &   69.55 & 1.64 &  16.92 & 1.64 &  27.59 & 1.52 &  65 \\
S3 & 2.21 & 466.50  & 2.20 & 1390.33 & 2.07 & 417.39 & 2.06 & 712.69 & 1.73 & 150 \\
\bottomrule
\end{tabular}

 \text{$\phantom{t}^{\text{a}}$} Experimental values from Ref.~\citenum{makarov2008two}.\hfill
\end{table}

Both QM/MM approaches outperform the vacuum and QM/PCM models in
reproducing the experimental relative differences between the vertical
transition energies.
As already pointed out in case of OPA spectra (see Figure \ref{fig:OPA-methods}
and Table \ref{tab:OPA-methods}), this is particularly evident in the case of
the S1-S3
difference. The relative intensity of the S1 and S3 transitions is correctly
reproduced by all the different methods, with the best agreement with experiment
shown by QM/TIP3P and QM/FQ.

The largest discrepancy is reported in case of the S2 transition for both relative
energies and intensities. The data reported in Table \ref{tab:TPA-methods} show
that the S1-S2 energy difference is indeed wrongly reproduced even by QM/MM
approaches ($\sim$ 0.2 eV vs. 0.38 eV in the experiment). Some differences
between the considered methods are encountered in case of relative intensities
between S1 and S2. In particular, QM/PCM is almost able to reproduce the
experimental intensity ratio between the two peaks (2.9 vs. 4.3), whereas
non-polarizable QM/TIP3P results are the worst (0.9 vs. 4.3). Such a huge
discrepancies probably reflects the fact that for this transition, HB
interactions play a minor role with respect to polarization effects.

Finally, the major effect of including polarization effects in the description
of solvent effects is the increase of the S3 TPA cross section, which is
reported by both QM/PCM and QM/FQ whereas QM/TIP3P is consistent with the vacuum
counterpart. Remarkably, in passing from a continuum to a discrete
approach (QM/FQ), the S3 TPA cross section decreases, thus moving closer to the
experimental value.

To conclude our comparison between computed and experimental TPA spectra of
\ac{R6G} in water, it is worth pointing out that experimental TPA measurements
are not a standard technique.\cite{makarov2008two} In fact, as reported in
Figure S5 in the SI, several experimental TPA spectra of \ac{R6G} in aqueous
solution have been reported previously in the literature,\cite{makarov2008two}
showing large discrepancies even in the measured spectra. Therefore, a
quantitative comparison is particularly challenging.

\section{Summary, Conclusions and Perspectives}\label{sec:conclusions}

In this paper, the extension of the QM/FQ approach to linear and quadratic response
in the quasienergy formalism has been presented for the first time. The
approach has been coupled to a classical MD simulation in order to have a
reliable sampling of the phase space. The computational protocol has been
applied to a challenging problem, \ie the OPA and TPA spectra of \ac{R6G} in
aqueous solution. Such a molecule is characterized by a transition (S2) that is
dark in OPA spectra due to symmetry selection rules, see Figure
\ref{fig:OPA-methods}. Thanks to the different selection rules, however, such a
transition is visible in TPA spectra, see Figure \ref{fig:TPA-methods}.

To analyze the importance of solvent effects for such a system, three different
approaches have been considered, from a continuum (QM/PCM) to two QM/MM approaches
(\ie{} nonpolarizable QM/TIP3P and polarizable QM/FQ).
Both QM/TIP3P and QM/FQ simulated OPA and TPA spectra show that the
inclusion of discrete water solvent molecules is essential to increase the
agreement between theory and experiment.
However, some discrepancies still remain and are principally related to the
OPA-dark S2 transition.
For this particular transition, polarization effects included by both QM/PCM and
QM/FQ seem to be crucial.
Although the agreement with experiment increases by including solvent
polarization, it still remains poor. This can be ascribed to several factors.
The choice of DFT as the quantum mechanical methodology has been recently reported to
systematically underestimate TPA cross sections if compared to reference
resolution-of-the-identity CC2 data.\cite{beerepoot2018benchmarking}
\ac{DFT} functionals are mostly tested on vertical excitation energies, and only
rarely on TPA cross sections, which for an OPA-dark transition is particularly
problematic.
In addition, as pointed out in Ref.~\citenum{milojevich2013surface},
the inclusion of vibronic coupling can play an important role in the
determination of TPA cross sections.\cite{macak2000electronic,macak2000simulations}
Finally, our polarizable QM/FQ approach models the interaction between the QM and MM portions in terms of electrostatic interactions only.
Some of the present authors have recently developed different approaches to
include non-electrostatic energy terms in QM/MM
approaches,\cite{giovannini2017disrep,curutchet2018density,giovannini2019epr} which can play a
relevant role in predicting both OPA and TPA spectra.

\begin{acknowledgement}

RDR thanks Filippo Lipparini (Università di Pisa) for the enlightening
discussions on the variational formulation of polarizable QM/classical models,
and Simen S. Reine (Hylleraas Centre for Quantum Molecular Sciences, University
of Oslo) for help in the parallel set up of the calculations presented in this
work.

RDR and LF acknowledge the support of the Research Council of Norway
through its Centres of Excellence scheme, project number 262695 and the
Norwegian Supercomputer Program through a grant for computer time (Grant No.
NN4654K).
RDR also acknowledges support by the Research Council of Norway through its
Mobility Grant scheme, project number 261873.

CC gratefully acknowledges the support of H2020-MSCA-ITN-2017 European Training Network
“Computational Spectroscopy In Natural sciences and Engineering” (COSINE),
grant number 765739.

\end{acknowledgement}

\begin{suppinfo}

Analysis of the convergence of the QM/FQ \ac{OPA} and \ac{TPA} spectra as a function of the number of snapshots.
Calculated \ac{OPA} and \ac{TPA} spectra in the whole energy range. Experimental TPA spectra.

Input and output files for the LSDALTON calculations reported in this work are available
on the Norwegian instance of the Dataverse data repository: \url{https://doi.org/10.18710/C9OZWV}.

\end{suppinfo}

{
\small
\bibliography{tpa-fq}

\begin{mcitethebibliography}{133}
\bibitem[Lin and Truhlar(2007)Lin, and Truhlar]{lin2007qm}
Lin,~H.; Truhlar,~D.~G. {QM/MM}: what have we learned, where are we, and where
  do we go from here? \emph{Theor. Chem. Acc.} \textbf{2007}, \emph{117},
  185--199\relax
\mciteBstWouldAddEndPuncttrue
\mciteSetBstMidEndSepPunct{\mcitedefaultmidpunct}
{\mcitedefaultendpunct}{\mcitedefaultseppunct}\relax
\EndOfBibitem
\end{mcitethebibliography}
}

\end{document}